\documentclass[conference, 10pt]{IEEEtran}

\usepackage{cite}
\usepackage{url}
\usepackage{graphicx}
\usepackage{color}
\usepackage{placeins}
\usepackage{float}
\usepackage{tabularx,colortbl}
\usepackage{ifthen}
\usepackage{bm}

\usepackage{amsmath,amssymb,amsfonts}

\makeatletter

\newcounter{author}
\renewcommand{\author}[2][]{
   \stepcounter{author}
   \@namedef{author@\theauthor}{#2}
   \@namedef{authorlabel@\theauthor}{#1}
}

\newcounter{address}
\newcommand{\address}[2][]{
   \stepcounter{address}
   \@namedef{address@\theaddress}{#2}
   \@namedef{addresslabel@\theaddress}{#1}
}

\newcommand{\alsep}{and}

\def\newmaketitle{\par%
  \begingroup%
  \normalfont%
  \def\thefootnote{}
  \def\footnotemark{}
  \let\@makefnmark\relax
  \footnotesize
  \footnotesep 0.7\baselineskip
  \normalsize%
  \twocolumn[\thenewmaketitle\@IEEEaftertitletext]%
  \if@IEEEusingpubid
     \enlargethispage{-\@IEEEpubidpullup}%
  \fi
  \endgroup
  \setcounter{footnote}{0}\let\maketitle\relax\let\@maketitle\relax
  \gdef\@thanks{}%
  \let\thanks\relax}

\def\thenewmaketitle{
  \newpage
  \begin{center}%
    \vskip0.2em{\Huge\@IEEEcompsoconly{\sffamily}\@IEEEcompsocconfonly{\normalfont\normalsize\vskip 2\@IEEEnormalsizeunitybaselineskip
   \bfseries\large}\@title\par}\vskip1.0em\par%
    \vspace{1ex}
    \newcounter{c@author}
    \newcounter{c@tmp}
    \ifthenelse{\value{author}=2}{%
      \newcommand{\liand}{ and }}{%
      \newcommand{\liand}{, and }}
    \ifthenelse{\value{address}<2}{%
      \@nameuse{author@1}%
      \stepcounter{c@author}%
      \whiledo{\value{c@author}<\value{author}}{%
        \setcounter{c@tmp}{\value{author}}%
        \addtocounter{c@tmp}{-\value{c@author}}%
        \ifthenelse{\value{c@tmp}=1}{%
          \renewcommand{\alsep}{\liand}}{\renewcommand{\alsep}{, }}%
        \stepcounter{c@author}\alsep \@nameuse{author@\thec@author}}\\%
    }
    {
      \@nameuse{author@1}${}^{(\ref{\@nameuse{authorlabel@1}})}$%
      \stepcounter{c@author}%
      \whiledo{\value{c@author}<\value{author}}{%
      \setcounter{c@tmp}{\value{author}}%
      \addtocounter{c@tmp}{-\value{c@author}}%
      \ifthenelse{\value{c@tmp}=1}{%
        \renewcommand{\alsep}{\liand}}{\renewcommand{\alsep}{, }}%
      \stepcounter{c@author}\alsep \@nameuse{author@\thec@author}%
        ${}^{(\ref{\@nameuse{authorlabel@\thec@author}})}$%
      }
    }
    \vspace{0.2ex}

    \ifthenelse{\value{address}>0}{%
      \ifthenelse{\value{address}=1}{
        {\@nameuse{address@1}}
      }
      {
        \newcounter{c@address}

        \begin{center}
        \whiledo{\value{c@address}<\value{address}}
        {
          \refstepcounter{c@address}
            ${}^{(\thec@address)}$\,%
              \label{\@nameuse{addresslabel@\thec@address}}%
              \@nameuse{address@\thec@address}\\ %
        }
        \end{center}
      } 
    }
    {
      \relax
    }
  \end{center}
}

\makeatother

\title{Loss-Optimized Reconfigurable Nonlocal Metasurface-aided Cavity Antenna}

\author[org1]{Minwoo Cho}
\author[org1]{Jeong-Hae Lee}
\author[org1]{Minseok Kim\textsuperscript{*}}

\address[org1]{School of Electronic and Electrical Engineering, Hongik University, 94 Wausan-ro, Mapo-gu, Seoul, 121-791, Korea \textsuperscript{*}minseok.kim@hongik.ac.kr}

\begin{document}

\newmaketitle

\begin{abstract}
This paper presents the design and experimental demonstration of a reconfigurable cavity-excited nonlocal metasurface antenna capable of wide-angle dynamic beam steering. The antenna is synthesized using a volume–surface integral equation (VSIE)–based framework that rigorously captures nonlocal mutual coupling among metasurface unit cells. To ensure physical consistency, the numerically characterized resistance–reactance ($R$–$X$) relationship of the tunable unit cells is directly incorporated into the synthesis, enabling precise far-field synthesis while minimizing Ohmic losses. The proposed approach is applied to a 10-GHz cavity-fed metasurface antenna composed of 24 independently controlled varactor-loaded unit cells. Numerical simulations and near-field measurements demonstrate stable beam steering up to $\pm40^\circ$ from broadside with excellent agreement between measured and simulated radiation patterns. These results confirm the effectiveness of the proposed framework for the realization of compact, reconfigurable cavity-excited metasurface antennas.
\end{abstract}


\section{Introduction}
\label{Introduction}
Metasurfaces are artificial thin-film structures composed of subwavelength unit-cell arrays designed to arbitrarily reshape scattered electromagnetic (EM) waves. Their macroscopic response is typically captured by representing the array as a homogenized, infinitesimal interface characterized by effective surface parameters (e.g., susceptibilities and impedances) which are rigorously linked to the surrounding fields through generalized sheet transition conditions (GSTCs)~\cite{IdemanIEEEPress}. Based on this framework, various wave-transformation techniques have been demonstrated, including anomalous reflection/refraction and polarization control~\cite{Ataloglou2021IEEEJM}.

To date, however, most conventional metasurfaces utilize the GSTCs from a strictly local perspective, which inherently restricts their functional scope. In this perspective, each unit cell is assumed to respond independently to the fields at its specific position. The validity of this local assumption is challenged when arbitrary field transformations are considered that do not necessarily satisfy the local power conservation condition (e.g., the conversion of a guided mode into a plane wave~\cite{Kim2021PRApplied}). Resolving the resulting local power mismatches requires power redistribution across the surface—a function that relies on mutual coupling between unit cells and is, by definition, a nonlocal process.

Several synthesis frameworks have been reported to capture these nonlocal processes. One approach leverages microwave network theory, utilizing commercial solvers to numerically extract the Green’s functions of constituent unit cells. While rigorous, this becomes computationally prohibitive as the electrical size of the structure increases. Alternatively, a more analytical design framework has been proposed that incorporates mutual coupling through volume surface integral equation (VSIE) formulations. Within this approach, the metasurface is modeled as an array of impedance strips representing homogenized unit cells. The required surface impedance values are then optimized by solving the VSIE via the method of moments (MoM) and are subsequently mapped to physical unit cells~\cite{Budhu2021IEEETAP}. Although this framework enables highly efficient designs, it has been largely restricted to idealized passive and lossless metasurfaces. This limitation arises because the unit-cell geometry is determined only after the synthesis is complete; consequently, the intrinsic correlation between resistance and reactance of a unit cell remains unknown during the optimization process. Nevertheless, recently, Ref.~\cite{Ataloglou2025TAP} reported a VSIE-based synthesis for reconfigurable intelligent surfaces (RIS) that directly incorporates the loss characteristics of tunable unit cells~\cite{Ataloglou2025TAP}. In their work, unit-cell geometries are pre-defined to identify the resistance-reactance ($R$-$X$) relationship of the unit cell, which is then enforced as a constraint during the optimization process. While this represents a significant step toward loss-aware synthesis, its application still remains centered on free-space wave transformation. Moreover, the iterative nature of MoM-based optimization remains a challenge, as it necessitates the repeated inversion of large matrices, leading to substantial computational cost.

Our previous work addresses some of these challenges by combining the idea of baffles with a VSIE-based formulation~\cite{Kim2025PRApplied}. This approach not only significantly reduces the computational costs associated with electrically large structures but also provides a route to directly transform internal cavity modes into desired radiation patterns, thereby facilitating the design of highly compact beam-forming platforms. However, it still remains limited to passive and lossless designs, leaving the physical loss of the tunable elements unaccounted for.

Building upon this foundation, the present work develops a synthesis framework specifically tailored to cavity-fed metasurfaces that explicitly accounts for dissipative effects of tunable elements. While adopting a similar strategy of pre-defining physical unit cells prior to VSIE-based synthesis, as reported in recent RIS designs~\cite{Ataloglou2025TAP}, the proposed approach is distinguished by its focus on dynamic wave transformation within cavity-excited architectures, where radiation arises from spatially varying leakage of internal modes. By directly incorporating the physical $R$–$X$ correlation into the VSIE formulation, the bias voltage distribution is optimized to jointly control radiation patterns and Ohmic dissipation. The proposed method is validated through full-wave simulations and near-field measurements of a fabricated prototype, demonstrating dynamic beam steering up to $\pm40^\circ$ from broadside.


\section{Proposed Architecture and Unit-Cell Characterization}
\label{section:Architecture and Homogenization}

\begin{figure}[!t]
\centering
  \includegraphics[width=0.9\columnwidth]{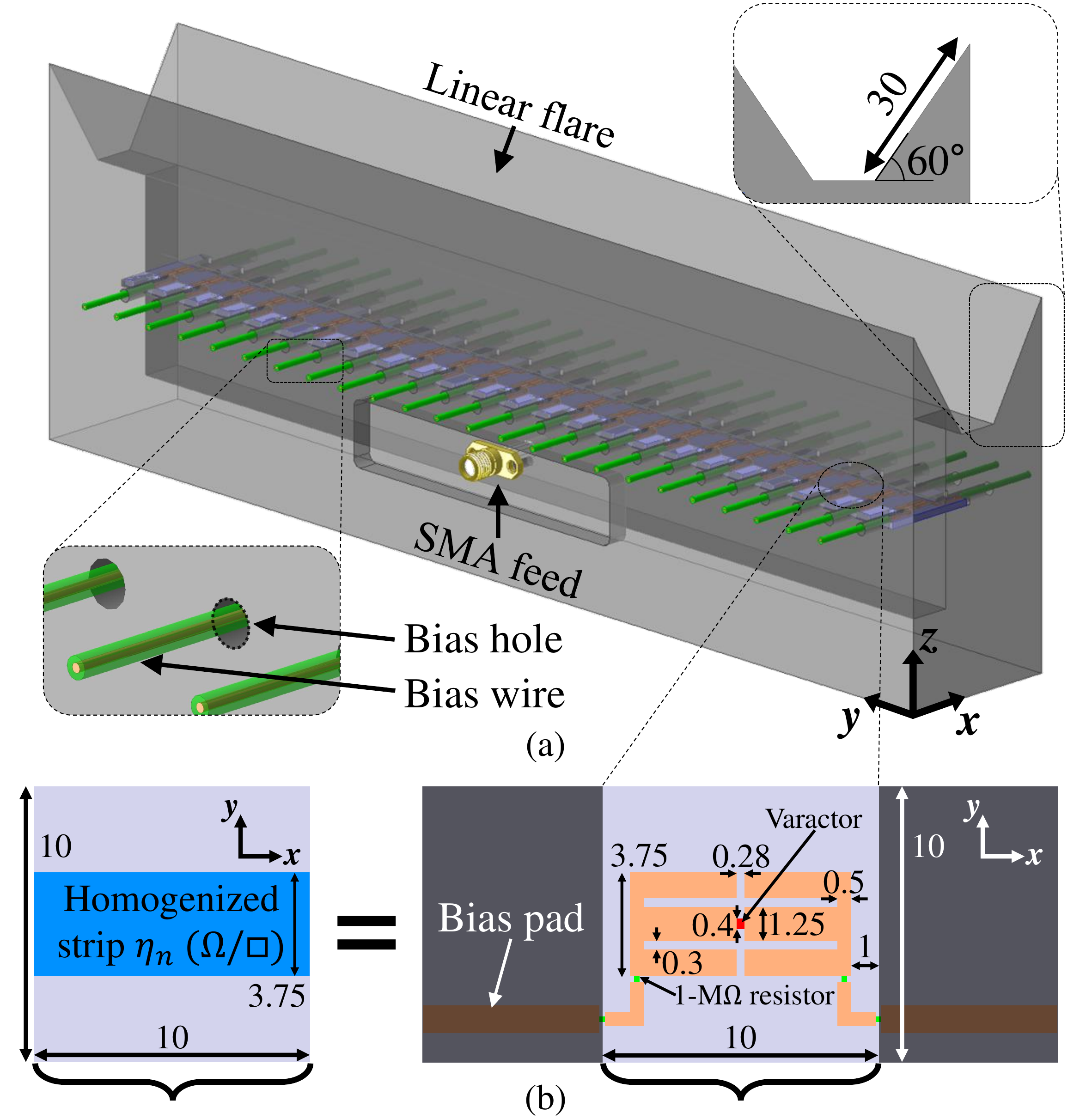}
  \caption{(a) Three-dimensional view of the proposed tunable cavity-excited metasurface antenna operates at 10~GHz ($\lambda_0\approx30$~mm), comprising 24 varactor-loaded unit cells with individually connected bias lines for independent voltage control. (b) Simplified numerical model, where the physical unit cell is represented by an impedance strip with a surface impedance of $\eta_n$. All dimensions are specified in millimeters (mm).}
  \label{fig:overall}
  \vspace{-0.5cm}
\end{figure}

Fig.~\ref{fig:overall}(a) depicts the proposed cavity antenna, which employs a nonlocal tunable metasurface as its radiating aperture. The term ``nonlocal'' reflects the design objective of utilizing inter-element mutual coupling to efficiently transform internal cavity modes into prescribed radiation patterns. The system is excited at 10 GHz ($\lambda_0 \approx 30$ mm) by an $x$-polarized $\text{TE}$ wave launched from a coaxial SMA feed to establish the cavity mode. Once the metasurface transforms this internal field, a surrounding linear flare guides the resulting radiation to enhance directivity.

The metasurface comprises 24 tunable unit cells, each integrated with a varactor diode from MACOM (MAVR-011020-14110) offering a variable capacitance range of 0.25--0.04 pF. These cells are arranged with a periodicity of $\lambda_0/3$, spanning a total aperture length of $8\lambda_0$. Detailed physical geometries are provided in Fig.~\ref{fig:overall}(b). DC bias is supplied via wires soldered to dedicated bias pads and routed through 3-mm diameter holes in the cavity walls. The bias pads are intentionally offset from the unit-cell centers to mitigate field leakage, and 1-M$\Omega$ chip resistors are incorporated to further suppress RF interference.

\begin{figure}[!t]
\centering
  \includegraphics[width=0.9\columnwidth]{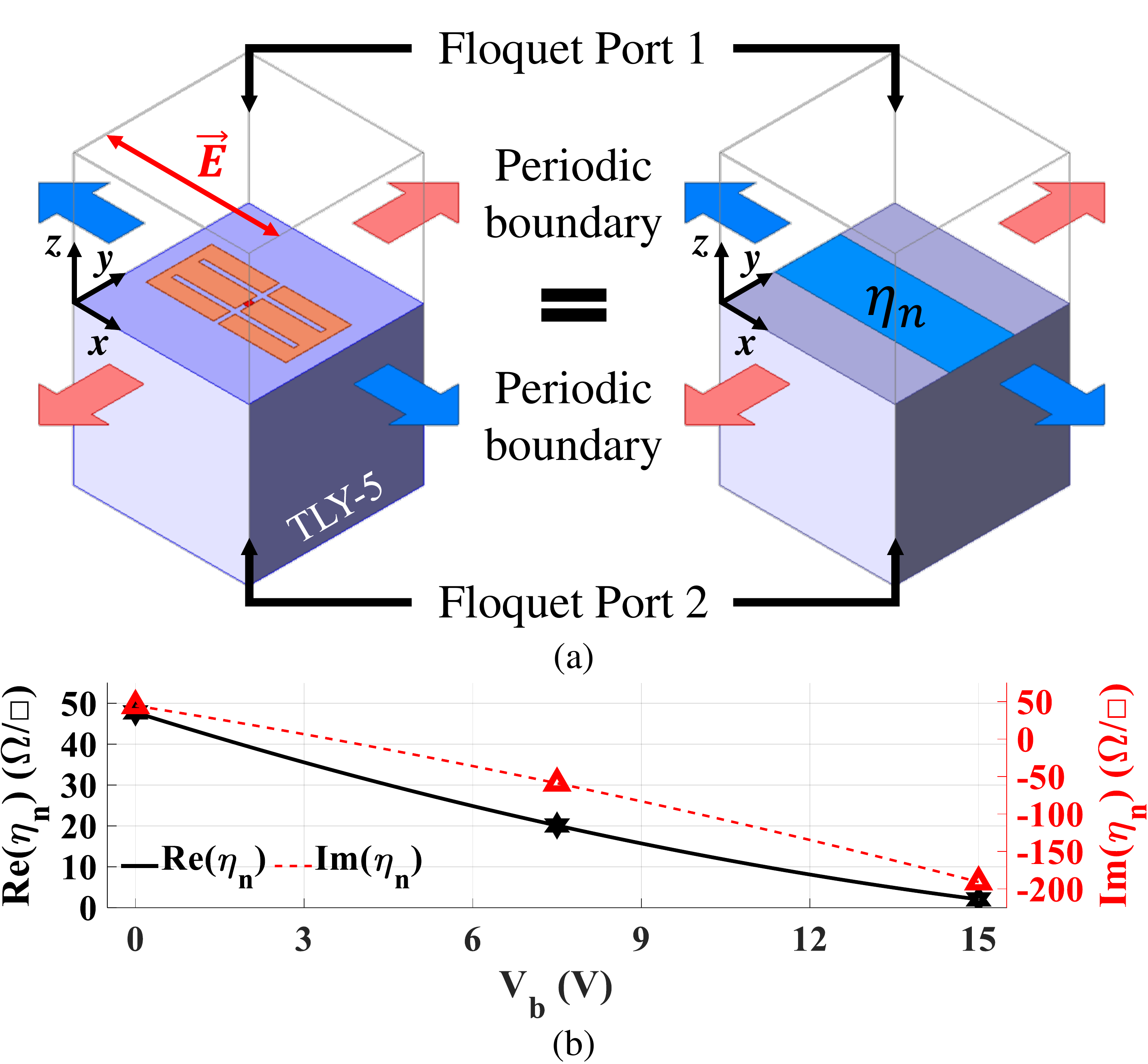}
    \caption{(a) Periodic boundary setup for matching a physical varactor-loaded unit cell to a homogenized complex impedance strip \(\eta_n\). (b) Mapping of $\eta_n$ as a function of $\mathbf{V}_\mathrm{b}$ via second-order polynomial interpolation.}
  \label{fig:eta_n}
  \vspace{-0.5cm}
\end{figure}

For efficient synthesis of the antenna via VSIE, the unit cells are modeled as homogenized impedance strips, as illustrated in Fig.~\ref{fig:overall}(b). Each strip's width, $w=3.75$ mm, is set equal to that of the physical unit cell and is characterized by a complex surface impedance, $\eta_n = r_n + jx_n$ $[\Omega/\square]$. The relationship between the physical unit cell and the homogenized model is established through a matching procedure shown in Fig.~\ref{fig:eta_n}(a). Specifically, the physical unit cell at a certain bias voltage ($V_{\text{b}}$) is first simulated under periodic boundary conditions in \texttt{ANSYS HFSS}, and the corresponding transfer impedance, $Z_{12}^{\text{phys}}$, defined as the voltage induced at Floquet port 1 by a unit current excitation at Floquet port 2, is extracted. Under an identical setup, the transfer impedance of the impedance strip, $Z_{12}^{\text{strip}}$, is computed by varying the strip resistance $r_n$ and reactance $x_n$. The bias voltage and $\eta_n$ are then mapped by matching $Z_{12}^{\text{phys}}$ with $Z_{12}^{\text{strip}}$, thereby ensuring that the homogenized model accurately captures the scattering response of the predefined physical structure. This matching is conducted at discrete bias voltages of 0, 7.5, and 15 V, after which a second-order polynomial interpolation is applied to define $\eta_n$ as a continuous function of $V_{\text{b}}$. The resulting interpolation and the corresponding fit, expressed in \eqref{eq:Eta_Vb_fit}, are shown in Fig.~\ref{fig:eta_n}(b).
\begin{subequations}
    \begin{align}
        &\text{Re}\left(\eta_n\left(V_b\right)\right) = 0.0852V_b^2-4.3267V_b+47.76\\
        &\text{Im}\left(\eta_n\left(V_b\right)\right) = -0.2516V_b^2-11.9133V_b+44.50
    \end{align}
    \label{eq:Eta_Vb_fit}
\end{subequations}


\section{VSIE-based Design Framework}
\label{sec:IE}

Having established the voltage-to-impedance mapping in \eqref{eq:Eta_Vb_fit}, the bias voltages required to transform a cavity mode into a prescribed radiation pattern are determined using the VSIE-based formulation. This framework, which precisely captures the effects of mutual coupling, begins by relating the total field to the induced currents throughout the antenna structure. Specifically, by denoting the induced currents along the metasurface, cavity walls, and substrate as, $\bm{I}_{\mathrm{c}}$, $\bm{I}_{\mathrm{d}}$, and $\bm{I}_{\mathrm{v}}$, respectively, the total field is expressed as,
\begin{equation}
\begingroup \setlength\arraycolsep{2pt}
\begin{bmatrix}
    E_t(\bm{\rho}_{\mathrm{c}}) \\ 
    E_t(\bm{\rho}_{\mathrm{d}}) \\ 
    E_t(\bm{\rho}_{\mathrm{v}})
\end{bmatrix}
=
\begin{bmatrix}
    \eta_n(\bm{\rho}_{\mathrm{c}}) & 0 & 0 \\
    0 & 0 & 0 \\
    0 & 0 & P(\bm{\rho}_{\mathrm{v}})
\end{bmatrix} 
\begin{bmatrix}
    \bm{I}_{\mathrm{c}} \\
    \bm{I}_{\mathrm{d}} \\
    \bm{I}_{\mathrm{v}}
\end{bmatrix},
\endgroup
\label{eq:Etot_mat}
\end{equation}
where $\bm{\rho}_k~(k \in \{\mathrm{c,d,v}\})$ denotes the observation locations, and $P(\bm{\rho}_{\mathrm{v}}) = [j\omega(\epsilon_r-1)\epsilon_0]^{-1}$ is the volumetric polarization coefficient, with $\epsilon_r=2.2$ for the TLY-5 substrate. 

Eq.~\eqref{eq:Etot_mat} shows that the unknown currents can be analytically determined, provided the total field and the surface impedances of the strips are defined. Since the latter is already established through the voltage-to-impedance mapping in Section II, the problem reduces to determining the total field. This field is composed of incident and scattered components, where the former is defined as a cylindrical wave excited by the coaxial SMA feed. The scattered field, in turn, is expressed as:
\begin{equation}
\begingroup 
\setlength\arraycolsep{2pt}
 \begin{bmatrix}
    E_s^\mathrm{c} \\ 
    E_s^\mathrm{d} \\ 
    E_s^\mathrm{v}
 \end{bmatrix}
 = 
 \begin{bmatrix}
    \mathbf{G}_{\mathrm{cc}} & \mathbf{G}_{\mathrm{cd}} & \mathbf{G}_{\mathrm{cv}} \\
    \mathbf{G}_{\mathrm{dc}} & \mathbf{G}_{\mathrm{dd}} & \mathbf{G}_{\mathrm{dv}} \\
    \mathbf{G}_{\mathrm{vc}} & \mathbf{G}_{\mathrm{vd}} & \mathbf{G}_{\mathrm{vv}}
 \end{bmatrix}
 \begin{bmatrix}
    \bm{I}_{\mathrm{c}} \\
    \bm{I}_{\mathrm{d}} \\
    \bm{I}_{\mathrm{v}}
 \end{bmatrix},
\endgroup
\label{eq:Esc_mat}
\end{equation}
where the entries of $\mathbf{G}_{pq}$ are evaluated using the two-dimensional free-space Green's function with pulse basis functions and point matching. Singular self-terms are handled using standard regularization based on small-argument approximations~\cite{Kim2025PRApplied}.

Substituting Eq.~\eqref{eq:Esc_mat} into Eq.~\eqref{eq:Etot_mat} then yields the solution for the unknown currents as,
\begin{equation}
\begingroup
\setlength\arraycolsep{2pt}
\begin{bmatrix}
 \mathbf{I}_{\text{c}} \\
 \mathbf{I}_{\text{d}} \\
 \mathbf{I}_{\text{v}}
\end{bmatrix}
\!\! = \!\!
\left(
 \begin{bmatrix}
 {\eta_n(\bm{\rho}_{\text{c}})} & {0} & {0}
 \\ {0} & {0} & {0}
 \\ {0} & {0} & {P}(\bm{\rho}_{\text{v}})
\end{bmatrix}
-
 \begin{bmatrix}
    \mathbf{G}_{\mathrm{cc}} & \mathbf{G}_{\mathrm{cd}} & \mathbf{G}_{\mathrm{cv}} \\
    \mathbf{G}_{\mathrm{dc}} & \mathbf{G}_{\mathrm{dd}} & \mathbf{G}_{\mathrm{dv}} \\
    \mathbf{G}_{\mathrm{vc}} & \mathbf{G}_{\mathrm{vd}} & \mathbf{G}_{\mathrm{vv}}
\end{bmatrix}
\right)^{\!\!-1\!\!}
 \begin{bmatrix}
    E_i^\text{c}
 \\ E_i^\text{d}
 \\ E_i^\text{v}
\end{bmatrix}.
\endgroup
\label{eq:I}
\end{equation}
As shown in Eq.~\eqref{eq:I}, the currents can be calculated as a function of the surface impedance $\eta_n$, which in turn depends on the bias voltage, $V_{\text{b}}$, according to the mapping in Eq.~\eqref{eq:Eta_Vb_fit}. By treating $V_{\text{b}}$ as the design variables, the current distribution is optimized to (i) minimize Ohmic losses while (ii) matching the far-field radiation to a prescribed pattern. Specifically, the total Ohmic loss is estimated by invoking Ohm's law as
\begin{equation}
P_{\mathrm{ohmic}}^{\mathrm{VSIE}}=\frac{1}{2} \sum_{n=1}^{N_{\text{strip}}} \sum_{m=1}^{N_{\text{seg}}}r_n\left|I_{c, (n,m)}\right|^2 \Delta_{\mathrm{c}},\label{eq:Pohm_summary}\end{equation}
where $N_{\text{strip}}=24$ is the number of impedance strips, $N_{\text{seg}}=5$ is the number of segments per strip, and $\Delta_{\text{c}}=0.75$ mm represents the segment width. Specifically, $I_{c, (n,m)}$ represents the induced current coefficient of the $m$-th segment within the $n$-th impedance strip. This segmental current is determined as a part of the unknown current vector $\mathbf{I}_{\text{c}}$ in Eq.~\eqref{eq:I}, where each strip is discretized into $N_{\text{seg}}$ pulse basis functions to accurately capture the current distribution along the metasurface. On the other hand, the relationship between the currents and the resulting far-field pattern is defined as,
\begin{equation}
\mathbf{E}_t^\text{ff}(\theta) = \mathbf{E}_i^\text{ff}(\theta) + \mathbf{G}^\text{ff}(\theta)\mathbf{I},
\label{eq:Eff_total_short}
\end{equation}
\noindent where $\mathbf{E}_i^\text{ff}(\theta)$ and $\mathbf{G}^\text{ff}(\theta)$ denote the incident far field and the far-field Green's function matrix, respectively~\cite{Kim2025PRApplied}.

Based on the formulations in Eq.~\eqref{eq:Pohm_summary} and Eq.~\eqref{eq:Eff_total_short}, particle swarm optimization (PSO) is employed to find the optimal set of bias voltages ($V_{\text{b}}$) that minimizes both the Ohmic loss and the deviation from the desired radiation pattern. By directly optimizing $V_{\text{b}}$, the resistance ($r_n$) and reactance ($x_n$) are uniquely determined for each unit cell through the mapping in Eq.~\eqref{eq:Eta_Vb_fit}. This approach ensures that the impedance values used during optimization are consistent with the predefined physical structure, effectively incorporating the complex loss characteristics into the synthesis framework. The following section numerically and experimentally demonstrates the feasibility of the proposed design framework.


\section{Numerical and Experimental Validation}
\label{sec:validation}
\begin{table}[b!]
\centering
\caption{Radiation Efficiency Comparison (VSIE vs. \texttt{HFSS})}
\renewcommand{\arraystretch}{1.50}
\begin{tabular}{c || c c c c c || c}
    \hline
        Steering Angle & $0^\circ$ & $10^\circ$ & $20^\circ$ & $30^\circ$ & $40^\circ$ & Average \\
    \hline
        $\text{e}_{\mathrm{rad}}^{\mathrm{VSIE}}$ & $0.73$ & $0.55$ & $0.66$ & $0.63$ & $0.64$ & $0.64$ \\
        $\text{e}_{\mathrm{rad}}^{\texttt{HFSS}}$ & $0.68$ & $0.61$ & $0.63$ & $0.65$ & $0.62$ & $0.63$ \\
    \hline
\end{tabular}
\label{tab:eta_rad_compare_db}
\end{table}
To validate the VSIE-based design framework presented in Section~\ref{sec:IE}, a prototype consisting of 24 unit cells is investigated. The prototype is first analyzed numerically using \texttt{ANSYS HFSS}, where the varactor diodes integrated into each unit cell are modeled as lumped RLC boundaries, as illustrated in Fig.~\ref{fig:overall}. The bias voltages are optimized following the procedure described in Section~\ref{sec:IE} to transform the cavity mode into directive beams at steering angles of $0^\circ$, $\pm10^\circ$, $\pm20^\circ$, $\pm30^\circ$, and $\pm40^\circ$. The optimized bias voltages are subsequently mapped to the corresponding resistance and reactance values of each unit cell using Eq.~\eqref{eq:Eta_Vb_fit}, and these parameters are encoded into the RLC boundary models. Additionally, an SMA connector is positioned at the center of the cavity, with its metallic pin placed at a distance of $\lambda_0/4$ from the cavity wall to minimize input reflection.

\begin{figure}[t!]
\centering
  \includegraphics[width=0.9\columnwidth]{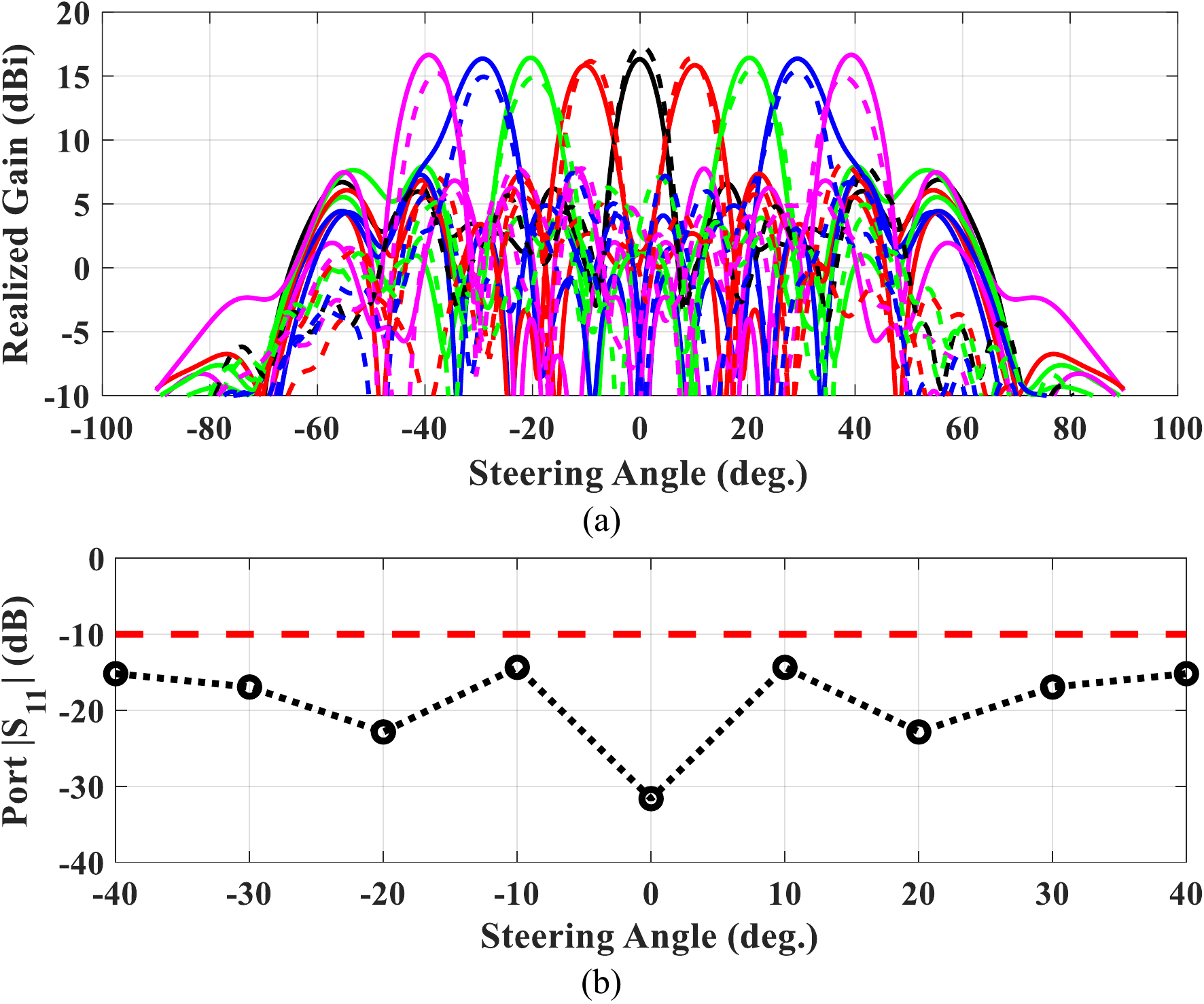}
  \caption{(a) Comparison between \texttt{HFSS} simulation results (solid lines) and near-field measurement results (dashed lines). (b) Measured input reflection coefficients for all steering angles, satisfying $|S_{11}| < -10$~dB.}
  \label{fig:OverlappedHFSSandMeasurement}
  \vspace{-0.5cm}
\end{figure}

The solid curves in Fig.~\ref{fig:OverlappedHFSSandMeasurement}(a) depict the simulated beam-steering performance.
The realized gain is calculated, clearly demonstrating that the radiated fields are steered toward the prescribed angles spanning from $-40^\circ$ to $+40^\circ$. Furthermore, Table~\ref{tab:eta_rad_compare_db} compares the radiation efficiencies predicted by the VSIE formulation with those obtained from \texttt{ANSYS HFSS} simulations. The close agreement between the two sets of results confirms the accuracy of the VSIE-based predictions and further validates the proposed design framework.

\begin{figure}[!t]
\centering
  \includegraphics[width=0.9\columnwidth]{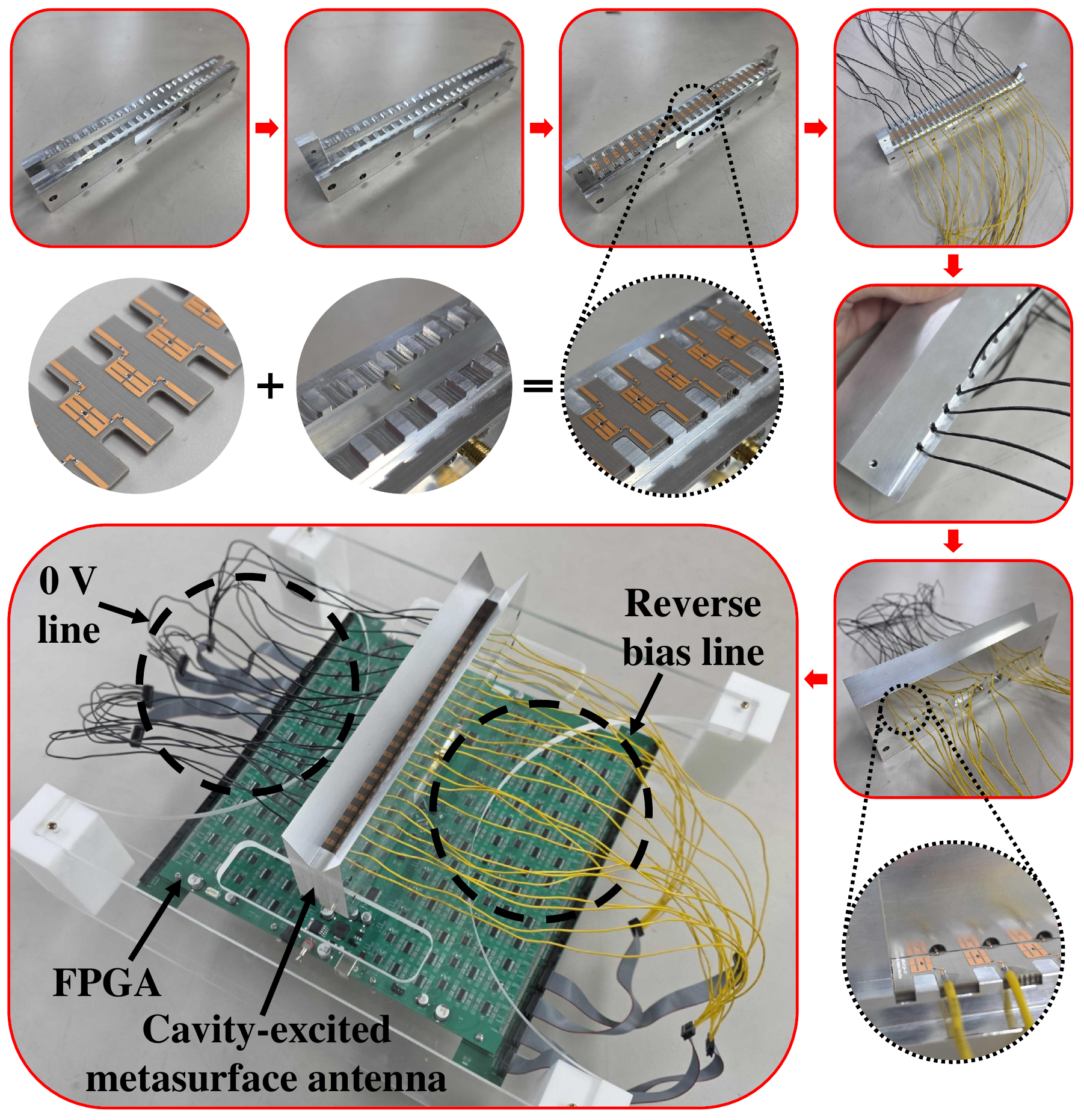}
  \caption{Fabrication and assembly sequence of the cavity-excited metasurface antenna, including CNC machining of the aluminum cavity, PCB fabrication of the varactor-loaded metasurface, and step-by-step assembly into the prototype.}
  \label{fig:fabrication}
\end{figure}

\begin{figure}[!t]
\centering
  \includegraphics[width=0.80\columnwidth]{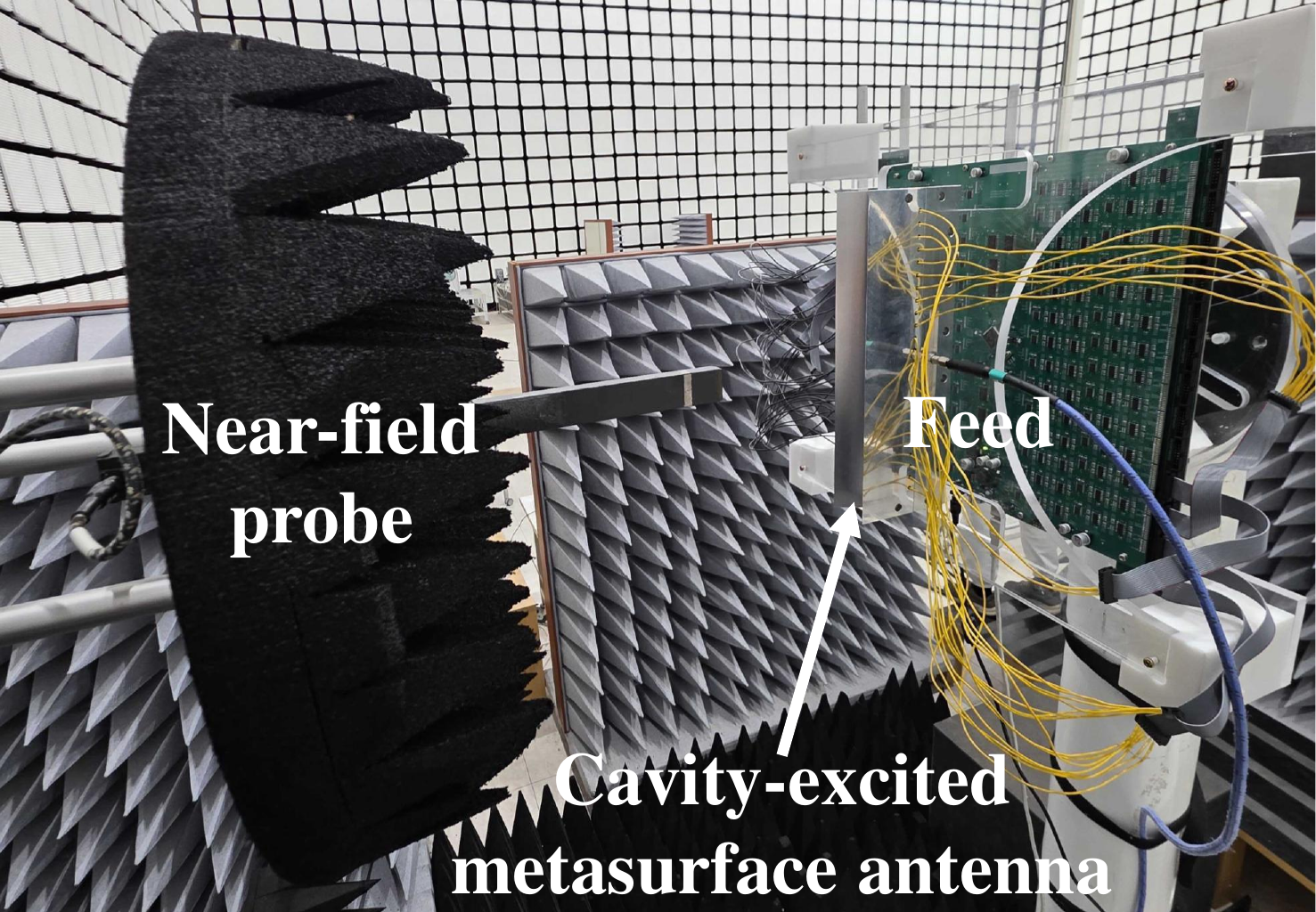}
  \caption{Near-field measurement setup.}
  \label{fig:NFMeasurementSetup}
  \vspace{-0.2cm}
\end{figure}

To experimentally substantiate these findings, the prototype is fabricated, with the assembly process illustrated in Fig.~\ref{fig:fabrication}. The metasurface layer is manufactured using a standard PCB fabrication process, while the cavity structure is realized through CNC machining to ensure mechanical robustness and dimensional accuracy.
Radiation measurements are conducted using the near-field measurement setup shown in Fig.~\ref{fig:NFMeasurementSetup}, and the measured near-field data are subsequently transformed into far-field radiation patterns.
The measured far-field results are overlaid with full-wave simulation results in Fig.~\ref{fig:OverlappedHFSSandMeasurement}(a), where the dashed curves represent the experimental data. Good agreement is observed between measurements and simulations across the entire steering range from $-40^\circ$ to $+40^\circ$. In addition, Fig.~\ref{fig:OverlappedHFSSandMeasurement}(b) presents the measured input reflection coefficient ($S_{11}$) for all steering angles, demonstrating that the antenna remains well matched with $|S_{11}| < -10$~dB throughout the entire scanning range. Overall, the strong agreement between experimental measurements and numerical predictions confirms that the proposed VSIE-based optimization framework is both robust and accurate, even when applied to practical, physically complex implementations.


\section{Conclusion}
This work establishes a design methodology for reconfigurable cavity-excited metasurface antennas and validates it through numerical and experimental demonstrations. The proposed approach employs a VSIE-based synthesis framework that rigorously accounts for nonlocal mutual coupling among unit cells while enforcing numerically characterized resistance--reactance ($R$--$X$) constraints of the tunable elements. By directly optimizing the bias voltage distribution within this formulation, the method enables systematic control of radiation patterns while minimizing Ohmic losses. Tailored to enclosed cavity-fed environments where radiation arises from spatially varying leakage of cavity modes, the framework provides a practical pathway for achieving wide-angle beam steering in a compact form factor.

\section*{ACKNOWLEDGEMENT}
This work was supported by the National Research Foundation of Korea (NRF) grants funded by the Korea government (MSIT) (RS-2024-00341191 and RS-2024-00343372).
\bibliographystyle{IEEEtran}
\bibliography{references/MinseokKim_Refs.bib}

\end{document}